\documentclass{gaa}
\usepackage{graphicx}
\usepackage{longtable}
\usepackage{supertabular}

\def\simgt{\ {\raise-.5ex\hbox{$\buildrel>\over\sim$}}\ }
\def\simlt{\ {\raise-.5ex\hbox{$\buildrel<\over\sim$}}\ }

\begin{document}

\title{Optical and Radio Variability of BL Lacertae}
\authorrunning{H. Gaur}
\titlerunning{Variability of BL Lac}
\author{Haritma Gaur,$^{1}$ Alok C.\ Gupta,$^{2,1}$ R. Bachev,$^{3}$ A. Strigachev,$^{3}$ E. Semkov,$^{3}$
 Paul J.\ Wiita,$^{4}$ A. E. Volvach,$^{5,6}$ Minfeng Gu,$^{1}$ A. Agarwal,$^{2}$  I. Agudo,$^{7}$
M. F. Aller,$^{8}$ H. D. Aller,$^{8}$
O. M. Kurtanidze,$^{9,10,11}$ S. O. Kurtanidze,$^{9}$ A. Lahteenmaki,$^{12,13}$ S. Peneva,$^{3}$  M. G. Nikolashvili,$^{9}$ L. A. Sigua,$^{9}$
 M. Tornikoski,$^{12}$ L. N. Volvach$^{5,6}$  }

   \institute{
 Key Laboratory for Research in Galaxies and Cosmology, Shanghai Astronomical Observatory,
Chinese Academy of Sciences, 80 Nandan Road, Shanghai 200030, China (haritma@shao.ac.cn)
\and
Aryabhatta Research Institute of Observational Sciences (ARIES), Manora Peak, Nainital -- 263002, India
\and
Institute of Astronomy and National Astronomical Observatory, Bulgarian Academy of Sciences, 
 72 Tsarigradsko Shosse Blvd., 1784 Sofia, Bulgaria 
\and
Department of Physics, The College of New Jersey, P.O.\ Box 7718, Ewing, NJ 08628-0718, USA 
\and
Radio Astronomy Laboratory of Crimean Astrophysical Observatory, Crimea
\and
 Taras Shevchenko National University of Kyiv, Kiev, Ukraine
\and
Instituto de Astrof\'{\i}sica de Andaluc\'{\i}a (CSIC),  
                 Apartado 3004, E-18080 Granada, Spain
\and
 Department of Astronomy, University of Michigan, 1085 South University Avenue,Ann Arbor, MI 48109, U.S.A.
\and
 Abastumani Observatory, Mt. Kanobili, 0301 Abastumani, Georgia  
\and
 ZAH, Landessternwarte Heidelberg, K$\ddot{o}$nigsthul 12, 69117 Heidelberg, Germany 
\and
 Engelhardt Astronomical Observatory, Kazan Federal University, Tatarstan, Russia
\and
 Aalto University Mets{\"a}hovi Radio Observatory, Finland
\and
 Aalto University Department of Radio Science and Engineering, Finland 
}


\abstract
{ We observed the prototype blazar, BL Lacertae, extensively in optical and radio bands during an active phase in the period 2010--2013  
when the source showed several prominent outbursts.  We searched for possible correlations 
and time lags between the optical and radio band flux variations using multifrequency data to learn about
the mechanisms producing variability.  
}
{During an active phase of  BL Lacertae, we searched for  possible correlations and time lags between multifrequency
light curves of several optical and radio bands. We tried to estimate any possible variability timescales and inter-band 
lags in these bands. }
{We performed optical observations in B, V, R and I bands from seven telescopes in Bulgaria, Georgia, Greece and India 
 and obtained radio data at 36.8, 22.2, 14.5, 8 and 4.8 GHz frequencies from three telescopes in Ukraine, Finland and USA. 
}
{ Significant cross-correlations between optical and radio bands are found in our observations with a delay of cm-fluxes with respect
to optical ones of $\sim$250 days. The optical and radio 
light curves do not show any significant 
timescales of variability. BL Lacertae  showed many optical `mini-flares' on short time-scales. Variations on longer term timescales are mildly chromatic
with superposition of many strong optical outbursts. In radio bands, the amplitude of variability is frequency dependent.
Flux variations at higher radio frequencies lead the lower 
frequencies by days or weeks.}
{  The optical variations are consistent with being dominated by a geometric scenario 
where a region of  emitting plasma moves along a helical path in a relativistic jet. The frequency dependence of the variability amplitude
supports an   origin of the observed variations intrinsic to the source. }

\keywords{galaxies: active -- galaxies: BL Lacertae objects: general -- galaxies: BL Lacertae objects:
individual: BL Lacertae -- galaxies: jets -- galaxies: quasars: general}

\maketitle

\section{Introduction}

BL Lacertae is the prototype of the BL Lac class of active galactic nuclei and has been observed in optical bands from
a long time. It is highly variable in all wavelengths ranging from radio to $\gamma$-ray bands (Raiteri et al.\ 2013
and references therein). This source is a favourite target of multi-wavelength campaigns of WEBT/GASP
and is well known for its intense optical variability on all accessible time-scales (Raiteri et al. 2009; Villata 
et al.\ 2009; Raiteri et al.\ 2010, 2011, 2013 and references therein). \\

Raiteri et al.\ (2009) presented the multi-wavelength data (from radio to X-rays)  of BL Lac 
of their 2007--2008 Whole Earth Blazar Telescope (WEBT) campaign and fitted the SEDs by an inhomogeneous, rotating helical jet model
(Villata \& Raiteri 1999; Ostorero et al.\ 2004; Raiteri et al.\ 2003)  which includes
synchrotron plus self-compton emission from a helical jet plus a thermal component from the accretion
disc. Larionov et al.\ (2010) studied the behaviour of BL Lacertae optical flux and colour variability
and suggested the variability to be mostly caused by changes of the Doppler factor.
Raiteri et al.\ (2010) studied the broad band emission and variability properties of the BL Lacertae during the period
2008--2009 and argued for a jet geometry where changes in our  viewing angle to the emitting regions plays an important role in the source's
multiwavelength behaviour.
Within the optical bands, a bluer when brighter chromatic trend was detected in BL Lacertae in previous studies
 (e.g., Villata et al.\ 2002; Gu et al.\ 2006; Gaur et al.\ 2012a; Agarwal \& Gupta 2015). \\ 

Connection of optical and radio light curves has shown significant correlations, with a
radio time delay of about 100 days, which can explained by the geometric effects
in a rotating helical path in a curved jet (Villata et al.\ 2009).
This source is known to vary on different timescales, with the usual abbreviations being IDV 
(intra-day variations -- within one night) to STV (short-term  variations -- on timescales of days to weeks) to LTV
(long-term  variations -- on timescales of months and years) (Gupta et al.\ 2004).  It
has been claimed to exhibit periodicities in its radio flux variations with a long
term component of P $\sim$8 years (Hagen-Thorn et al.\ 1997;
Villata et al.\ 2004; Villata et al.\ 2009). 
 Raiteri et al.\ (2013) collected exceptional optical sampling by the GLAST-AGILE Support Program of the
WEBT during the outburst period of the BL Lacertae (2008--2012) and performed cross-correlations between the 
optical--$\gamma$-ray and X-ray--mm wavelengths which suggest that the region producing the mm and X-ray radiation is located downstream
from the optical and $\gamma$-ray emitting zone in the jet. They found a significant cross-correlation between the optical and mm flux densities
with a time lag of 120-150 days.  Recently, Guo et al.\ (2015) analysed the historical light curves of optical
and radio bands from 1968--2014 to find possible periods of $\sim$1.26 and $\sim$7.50 years, respectively. \\

The search for correlations between different bands provides crutial information for the emission mechanism and 
since the radio and optical bands are both attibuted to the synchrotron emission from the relativistic electrons 
in the jets of blazars, they are of particular importance. A key motivation of this study is to investigate the 
correlated optical and radio variability (at cm wavelengths) of BL Lacertae during its
active state in 2010--2013 when the source  showed multiple outbursts in both of these bands. Also, we studied the nature of 
short- and long-term variability in optical bands as well as variability behaviour in radio bands and their possible variability
timescales.  Over the course of 3 years, we performed quasi-simultaneous optical 
multi-band photometric data from seven telescopes in Bulgaria, Greece, Georgia and India on 192 nights.
The radio data were observed from Ukraine, Finland and USA at five frequencies, 36.8, 22.2, 14.5, 8 and 4.8 GHz on 302 nights.\\

The paper is presented as follows: in section 2, we briefly describe the observations and data reductions. We present our results
in section 3 and section 4 includes a discussion and our conclusions. \\

\section{Observations and data reduction}
\subsection{Optical Data}
 Observations of BL Lacertae started on June 2010 and ran through July 2013 and the entire timeline
of the observational period, along with the long-term light curves are shown in Figure 1.
The observations were carried out at seven telescopes in Bulgaria, Greece, Georgia and India. 
The telescopes and cameras that were involved in these observations are described in detail in Gaur et al.\ (2012b).  \\

 Most of the observations were made at the 50/70 cm Schmidt and the 2m Ritchey--Chretien telescopes at
 Rozhen National Astronomical Observatory, Bulgaria, the 60 cm Cassegrain Telescope at Astronomical Observatory Belogradchik, 
Bulgaria, and the 1.3m Skinakas Observatory, Crete, Greece during the period June 2010 and July 2013.
Instrumental magnitudes and the comparison stars were extracted using the MIDAS package \texttt{DAOPHOT} with an aperture radii of 4 arcsec.
We used data of 50 days of observations from Rozhen 50/70 cm telescope, 35 from 2m Rozhen, 72 from Belogradchik and 25 
 from Skinakas Observatory. \\

The calibration of the source magnitude was performed with respect to the comparison stars B, C and H 
(Fiorucci \& Tosti 1996). The host galaxy of BL Lacertae is relatively very bright, hence, in order to remove its contribution from the
observed magnitudes, we first dereddened the data using the Galactic extinction coefficients of Schlegel
et al.\ (1998) and then subtracted the host galaxy contribution
from the observed magnitudes corresponding to that aperture radii (using Nilsson et al.\ 2007) in order to avoid its contamination in the
extraction of colour indicies. \\

 Observations of BL Lacertae on 10, 11, 12 and 14 June 2010 were carried out using the 1.04m Sampurnanand telescope located at Nainital,
India. Pre-processing of the raw data which includes bias subtraction, flat-fielding and cosmic ray removal was performed 
using standard data reduction procedures in IRAF. Reduction of the image frames was done using \texttt{DAOPHOT II}. Aperture photometry
was carried out using four concentric aperture radii, i.e., $\sim$ 1 $\times$ FWHM, 2 $\times$ FWHM, 3 $\times$ FWHM and 4 $\times$ FWHM.
We found that aperture radii of 2 $\times$ FWHM always provides the best S/N, so we adopted that aperture for our final results. \\

The observations at the Abastumani Observatory were conducted on 9, 11, 12, 14, 15
and 16 December 2010 at the 70-cm meniscus telescope (f/3). These measurements
were made with an Apogee CCD camera Ap6E (1K $\times$ 1K, 24 micron square pixels)
through a Cousins R filter with exposures of 60--120 sec.   Pre-processing of the raw data is done by IRAF. 
Reduction of the image frames were done using \texttt{DAOPHOT II}. An aperture radius of 5 arcsec was used
for data analysis. \\

Finally, we also include the published R band data from GASP/WEBT observations (Raiteri et al.\ 2013) during this 
period 2010--2013 where magnitudes are extracted for the BL Lac and the standard stars using an aperture radius 
of 8 arcsec. Hence, we dereddened the data using the Galactic extinction coefficient of Schlegel et al.\ (1998) 
and then removed the host galaxy contribution as described above. All the  optical data are presented in the top panel of Figure 1.

\subsection{Radio Data}

The observations were carried out with the 22-m radio telescope (RT-22) at the 
Crimean Astronomical Observatory (CrAO).
For our measurements, we used two similar Dicke switched radiometers of 22.2 
and 36.8 GHz. The antenna temperatures from sources were measured by the standard on--on
method. Before measuring the intensity, we determined the source position by
scanning. The radio telescope was then pointed at the source alternately by the principal
and reference (arbitrary)
beam lobes formed during beam modulation and having mutually orthogonal
polarizations. The antenna temperature from a source was defined as the
difference between the radiometer responses averaged over 30 s at two
different antenna positions. Depending on the intensity of the emission from
sources, we made a series of 6--20 measurements and then calculated the mean
signal intensity and estimated the rms error of the mean. \\

\begin{figure*}
   \centering
\includegraphics[width=18cm , angle=0]{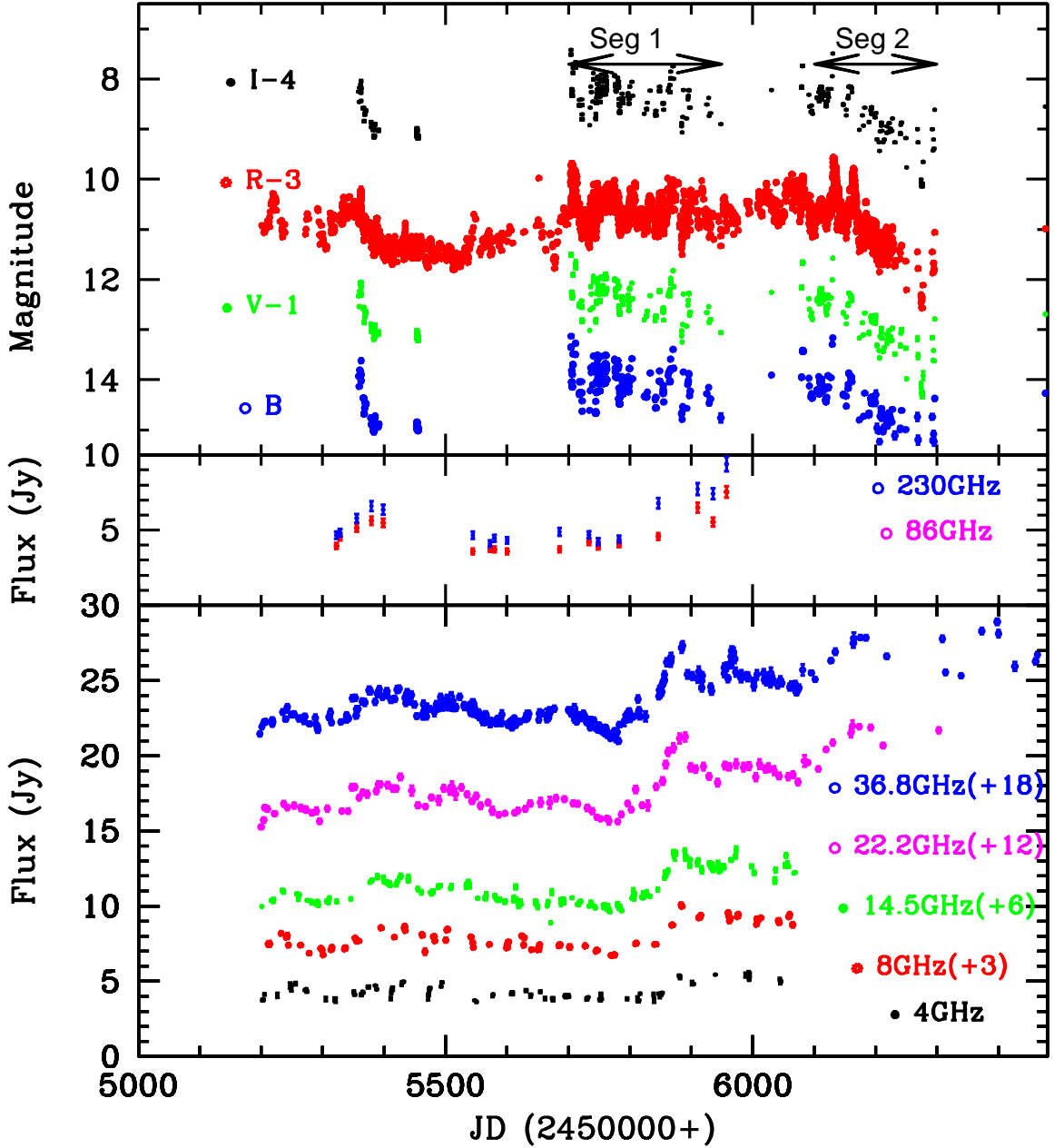}
\caption{The entire timeline of the observations in optical B, V, R and I bands (upper panel), the radio bands at 230 and
86 GHz (middle panel) and at 36.8 (and 37.0), 22.2, 14.5, 8.0 and 4.8 GHz frequencies (lower panel). The individual light curves 
and their offsets  are labeled.} 
 \end{figure*}
 
The gain of the receiver was monitored using a noise generator every 2 to 3 hours.
The orthogonal polarization of the lobes allowed us to measure the total
intensity of the emission from sources, irrespective of the polarization of
this emission. Absorption in the Earth's atmosphere was taken into account
by using atmospheric scans made every 3 to 4 hours. The errors of the calculated
optical depths are believed to be less than 10\%. 
 The errors of the measured flux densities include the uncertainties of: (1) the
detected mean value of the antenna temperature of the sources; (2) the
calibration source measurements; (3) the noise generator level measurement;
and (4) the atmosphere attenuation corrections.  The main contributions to
the quoted errors come from the first two terms. The flux density scale of observations was 
calibrated using DR 21, 3C 274, Jupiter and Saturn. \\

\begin{figure*}
   \centering
\includegraphics[width=5cm , angle=0]{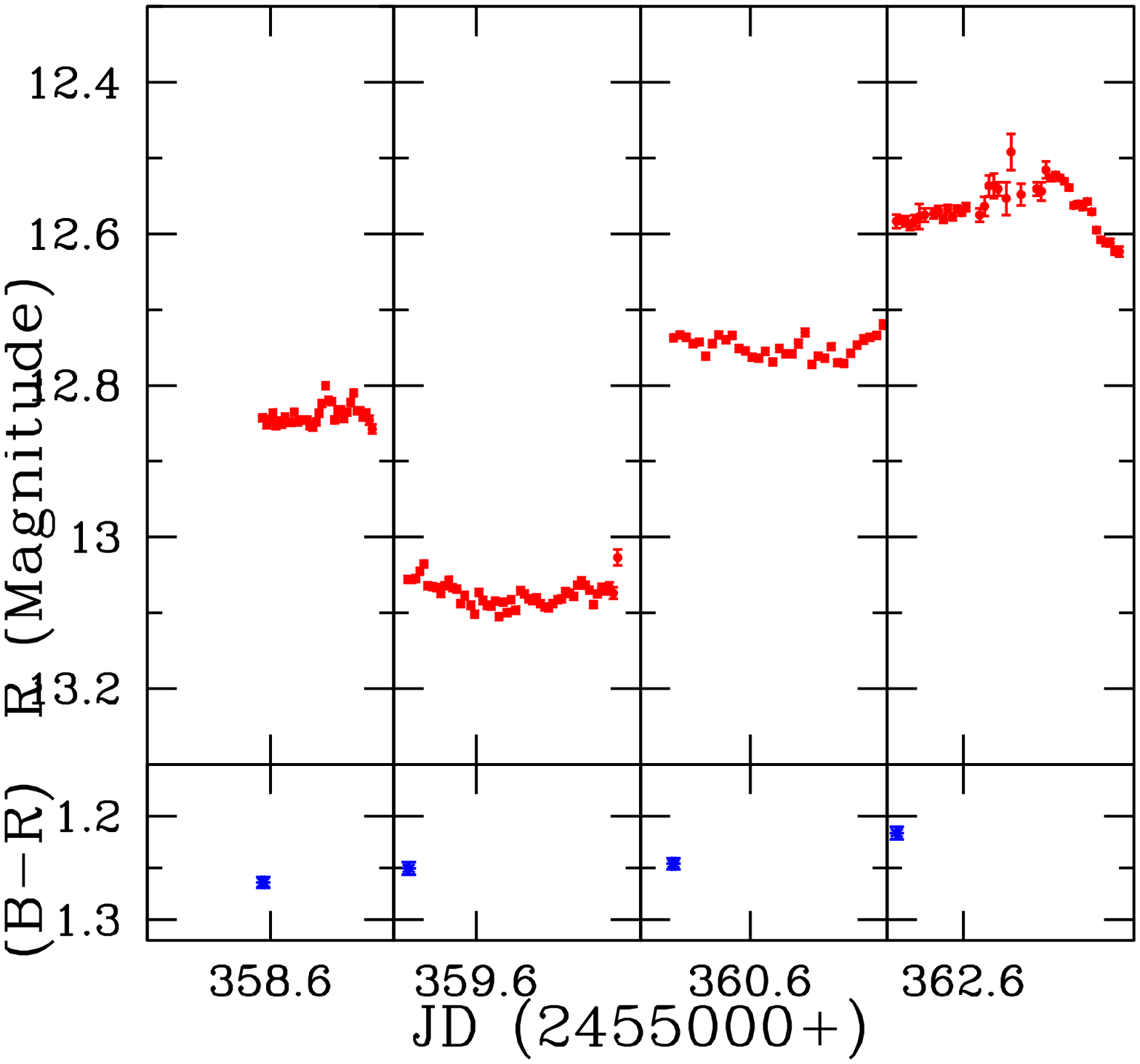}
\includegraphics[width=5cm , angle=0]{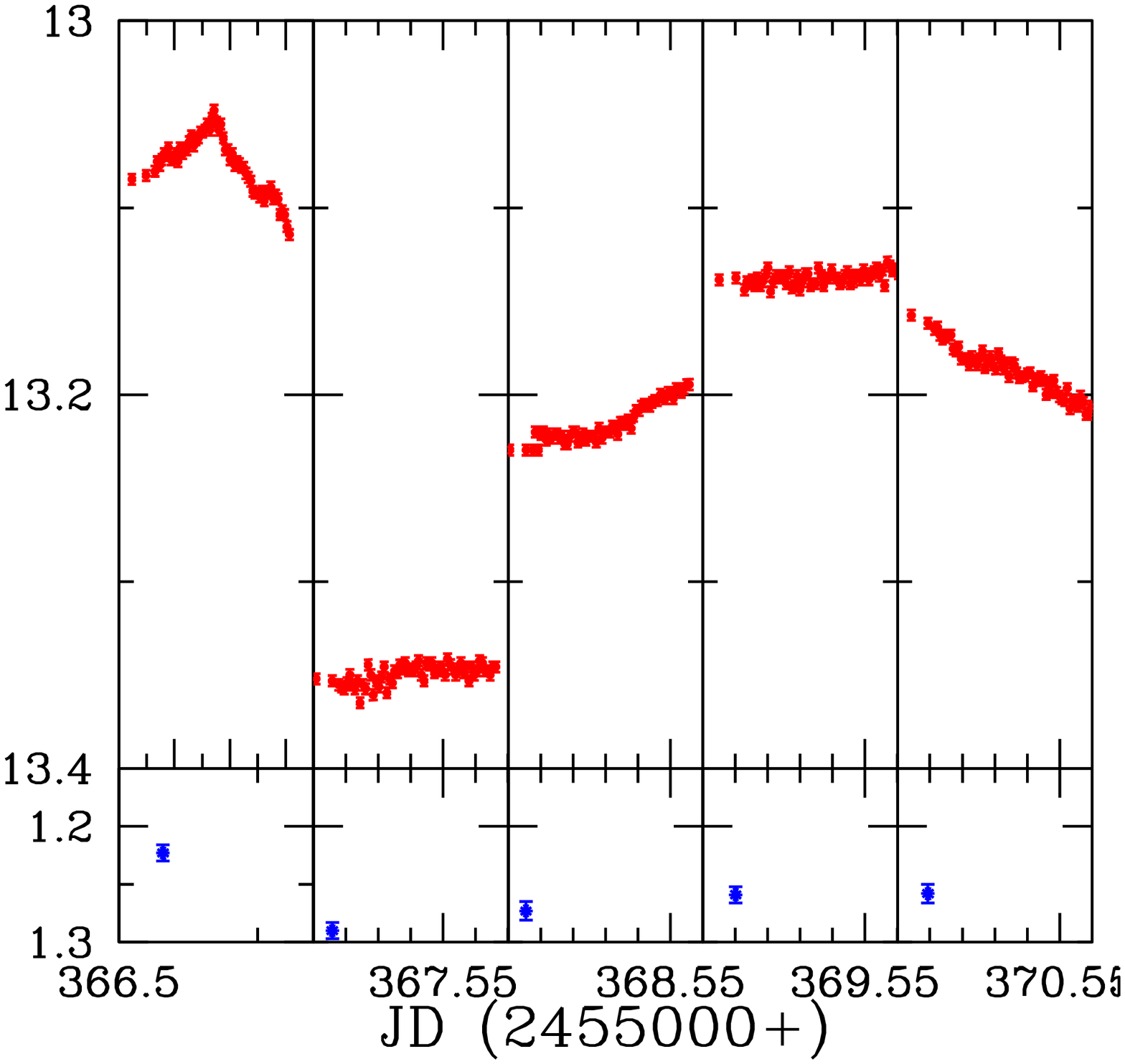}
\includegraphics[width=5cm , angle=0]{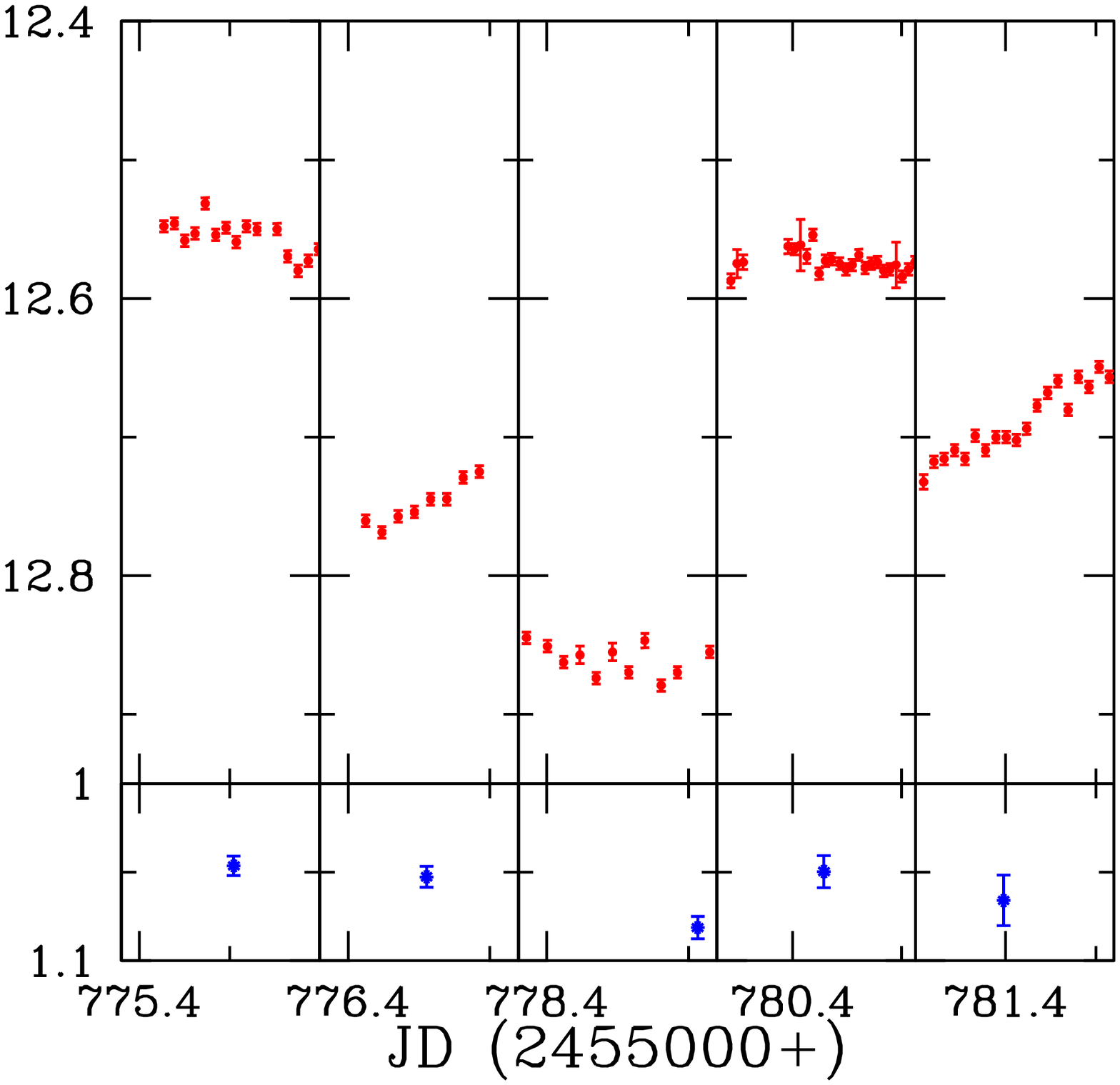}
\includegraphics[width=5cm , angle=0]{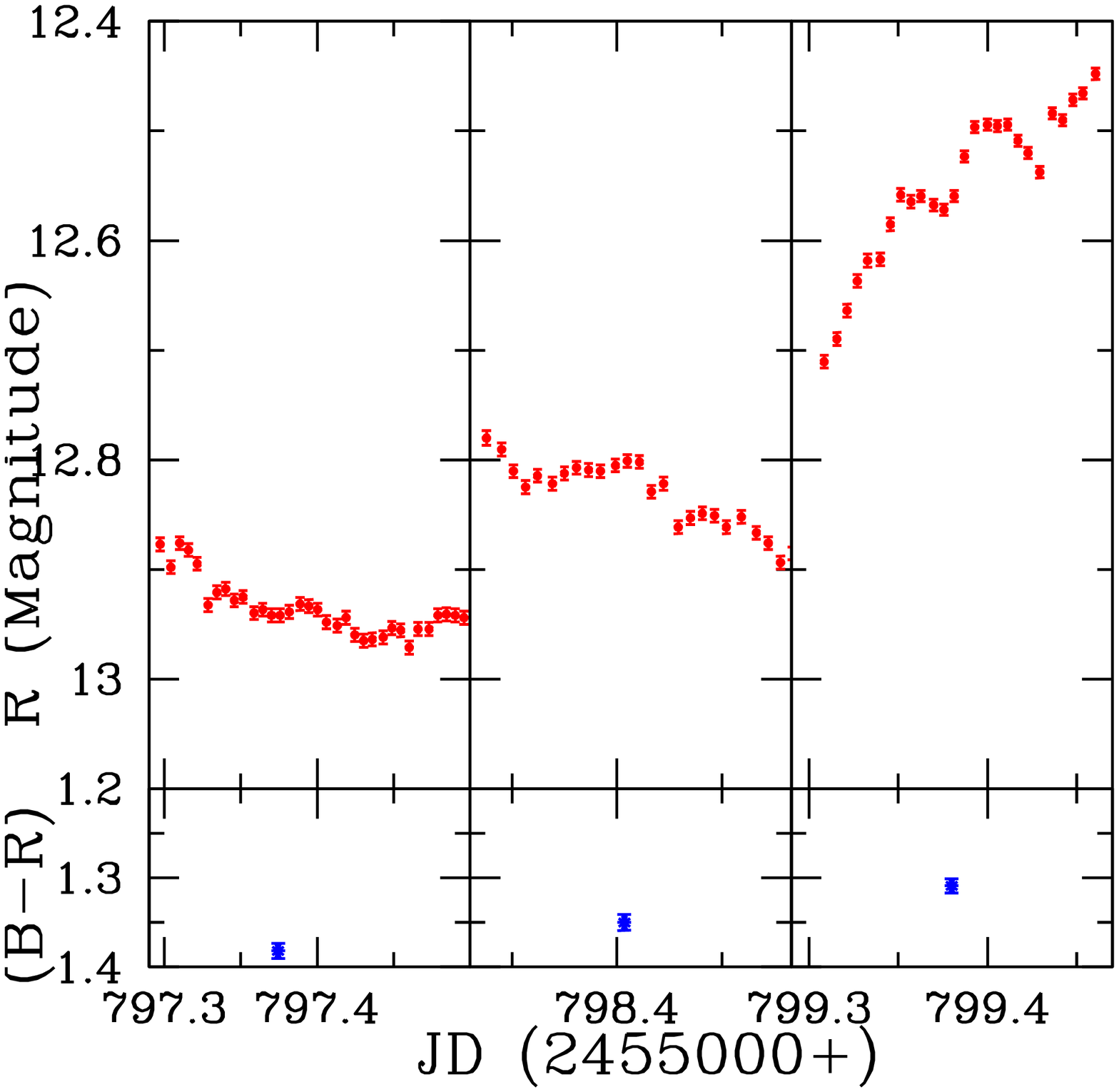}
\includegraphics[width=5cm , angle=0]{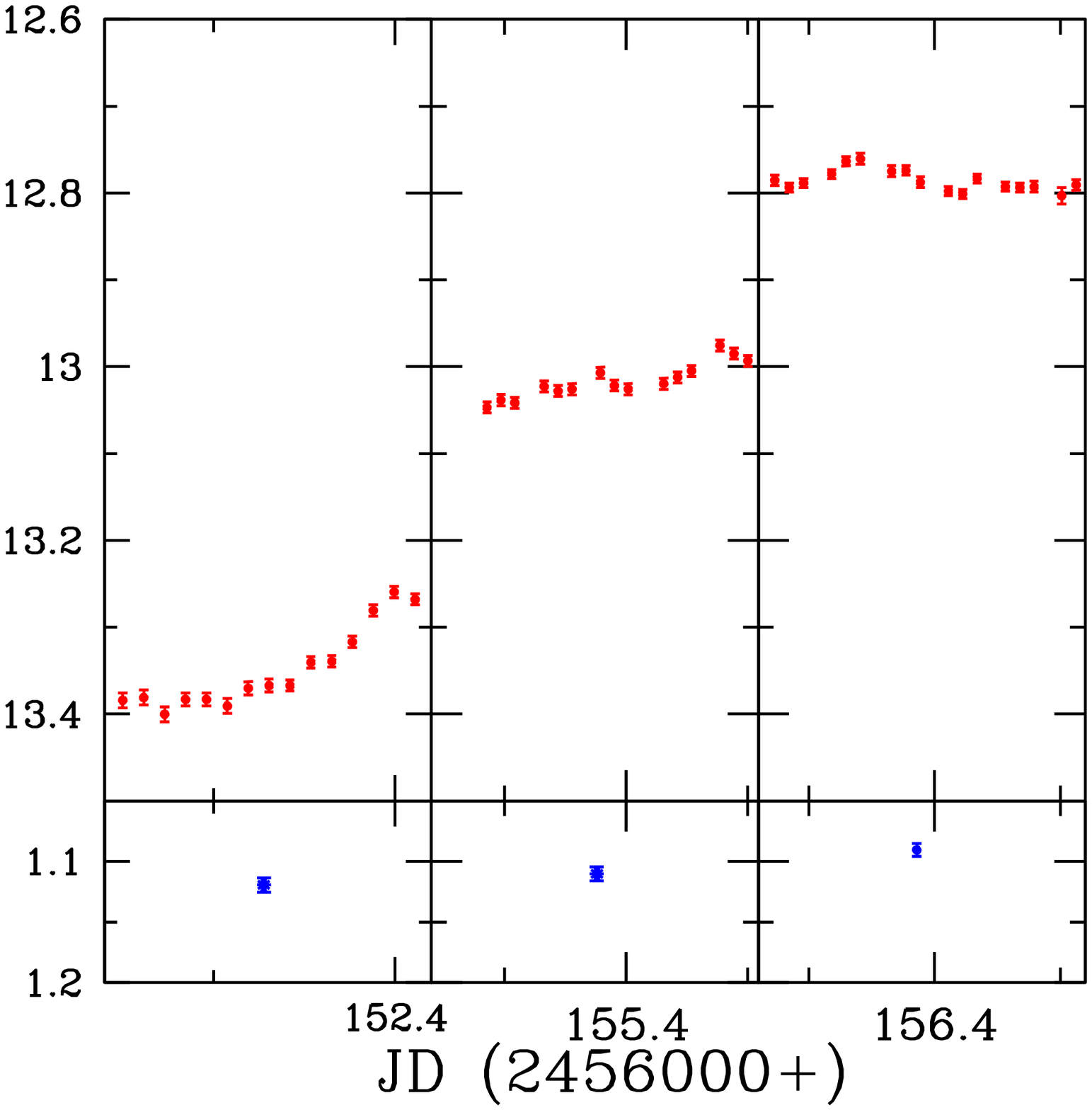}
\includegraphics[width=5cm , angle=0]{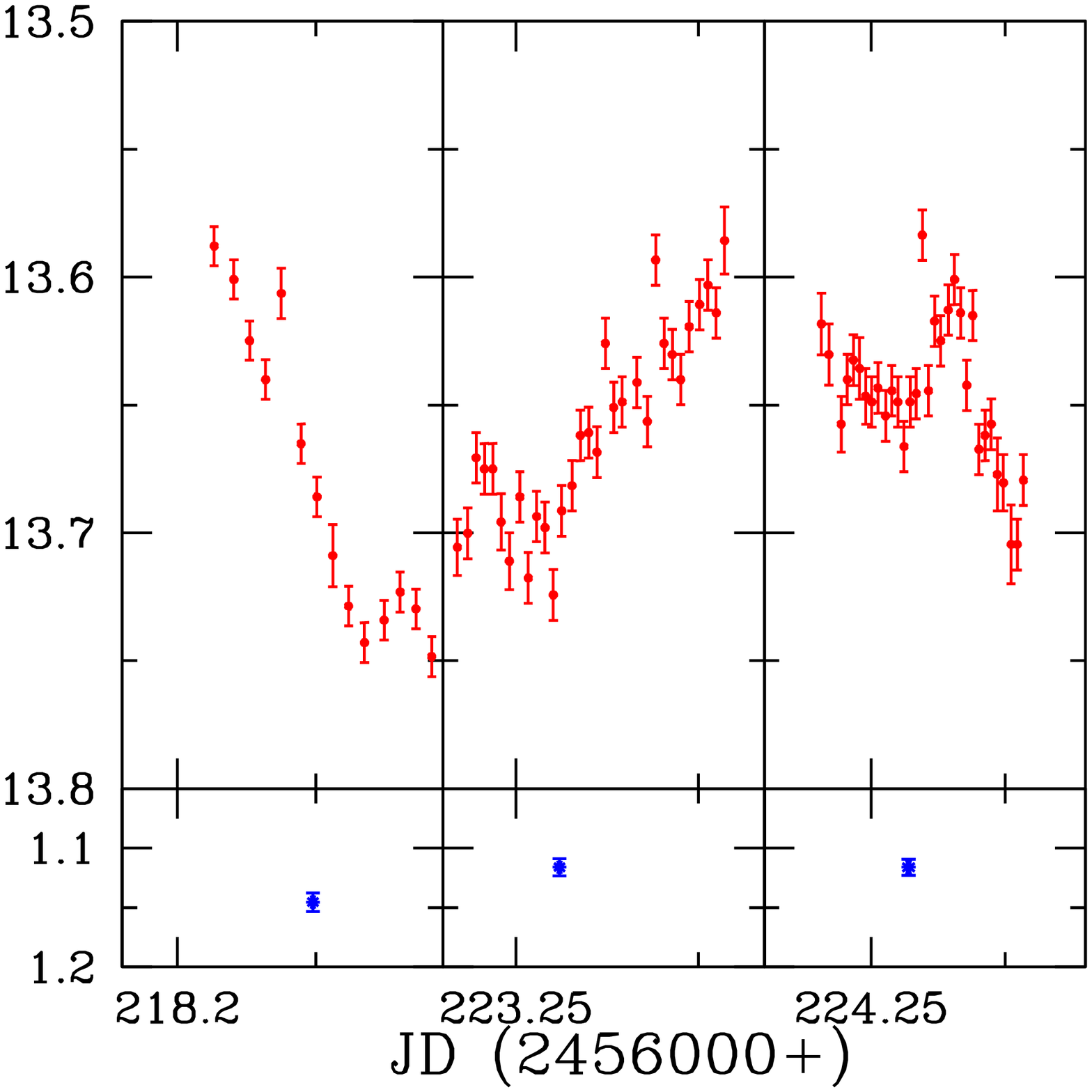}
\caption{ Short-term variability data for BL Lacertae in the R band (in upper panel portions) for each labeled date
and the respective average colour indices (B-R) for those nights (lower portions).}
 \end{figure*}
 
Observations at 37.0 GHz were made with the 14~m radio telescope of Aalto University's Mets\"ahovi Radio Observatory in Finland.
Data obtained at Mets\"ahovi and RT-22 were combined in a single array to supplement each other.
The flux density scale is set by observations of DR 21. Sources NGC 7027,
3C 274 and 3C 84 are used as secondary calibrators.
A detailed description of the data reduction and analysis of Mets\"ahovi data is given in Teraesranta et al.\ (1998).
The error estimate in the flux density includes the contribution from the
measurement rms and the uncertainty of the absolute calibration. \\

The lower frequency  flux density observations were obtained with the University of Michigan 26-m equatorially-mounted, prime focus, 
paraboloid (UMRAO) as part of the University of Michigan extragalactic variable source-monitoring program (Aller et al.\ 1985). 
Both total flux density and linear polarization observations were obtained as part of  the program’s measurements. 
Each daily-averaged observation of the target  consisted of a series of 8 to 16 individual measurements obtained over 
 25 to 40 minutes. At 14.5 GHz the polarimeter consisted of dual, rotating, linearly-polarized feed horns which were
 placed symmetrically about the paraboloid’s prime focus; these fed a broadband, uncooled HEMPT amplifier with a
 bandwidth of 1.68 GHz. At 8.0 GHz an uncooled, dual feed-horn beam-switching polarimeter and an on--on observing
 technique were employed; the bandwidth was 0.79 GHz. At 4.8 GHz a single feed-horn system with a central operating 
frequency of 4.80 GHz and a bandwidth of 0.68 GHz was used.  The adopted flux density scale is based on Barrs et al.\
 (1977) and uses Cassiopeia A (3C 461) as the primary standard.  In addition to the observations of this primary 
standard, observations of nearby secondary flux density calibrators selected from a grid were interleaved with the 
observations of the target source every 1.5 to 3 hours to verify the stability of the antenna gain and to verify the 
 accuracy of the telescope pointing. For observations of BL Lacertae, Cygnus A (3C 405),  DR 21 and NGC 7027 were used as calibrators. \\

We also included the published radio data at 230 and 86 GHz of Raiteri et al.\ (2013) from the IRAM 30m Telescope. 
The calibration procedure for the IRAM 30m Telescope's data was described in detail in Agudo et al.\ (2010, 2014).
The entireity of the radio band data are presented in the middle and the bottom panels of Figure 1.

\section{Results}

\subsection{Optical Flux and Colour Variations}
{\bf Short-term Variablity Timescales}
We first discuss  the variations that  occured on time-scales of days and weeks. These variations are seen by
combining the continuous observations and the observations which are done within each week. The light curves showing the short-term
variability are shown in the upper portions of each panel in Figure 2. It is obvious from the figures that during our
 observations BL Lacertae was very active and showed significant flux variations on short-term time-scales. Typical
 rates of brightness change were  0.2--0.3 mag/day. \\

We have examined the nature of colour variations along with the flux variations on short-term time-scales (shown in the bottom portions of
Figure 2 panels) and we notice that colour variations generally follow flux variations, in the sense that the
source is bluer when brighter.  \\

{\bf Long-term Variability Timescales}
We now discuss  the variations that occured on months-like time-scales. The long term variability light curves are 
shown in Figure 1 (upper panel). We found significant flux variations in the B, V, R and I bands and the variability in all the bands
appears to be well correlated. During the first full multi-colour observing season (segment 1), 
 a roughly contant baseline flux is
observed with small flares superimposed on it. In the next observing season (segment 2),
BL Lacertae showed a $\sim$2.5 magnitude decay leading to a decrease of the variability amplitude. \\

In order to examine the colour variability, we calculated (B-R) and (V-R) colour indices and the resulting
colour--magnitude diagrams are shown in Figure 3.
A linear fit is drawn in each plot and a slope of $0.039 \pm 0.013$ is found for (B-R) vs.\ R, with a linear
 Pearson correlation coefficient $r=0.224$ and its probability value, $p=0.003$.  The corresponding values for (V-R) vs.\ R are  
0.042 $\pm$ 0.005   with $r=0.512$ and $p=3.4 \times 10^{-13}$.   Together, these suggest 
a significant positive correlation between the colour index and brightness, in the sense of BL Lacertae being bluer-when-brighter.\\

On short-term timescales, the significant flux and colour variations can be explained by pure intrinsic phenomena,
such as shocks accelerating electrons in the turbulent plasma jets which then cool by synchrotron emission (e.g., Marscher 2014;
Calafut \& Wiita 2015) or the evolution of the electron density distribution of the relativistic particles leading to variable
synchrotron emission (e.g., Bachev et al. 2011). The faster variations are mostly associated with the spectral changes 
and are related to  very rapid electron injection and cooling processes. \\

Longer term variations are generally explained by a mixture of 
intrinsic and extrinsic mechanisms.  The former are often thought to involve a plasma blob moving through the helical structures 
of the magnetic field in the jets (e.g., Marscher et al.\ 2008),
which leads to  variable compression and polarization (Marscher et al.\ 2008; Raiteri et al.\ 2013; Gaur et al.\ 2014) or shocks in the helical jets
(e.g.\ Larionov et al.\ 2013).  The latter include the geometrical effects where our changing viewing angle to a moving, discrete emitting region
  causes variable Doppler boosting of the emitting radiation (e.g\ Villata et al.\ 2009; Larionov et al.\ 2010; Raiteri et al.\ 2013). 
Hence, the long term behaviour in blazar light curves is likely due to the superposition of both mechanisms (e.g.\ Pollack et al.\ 2015).
 The long term trend, which is only mildly chromatic and may be quasi-periodic (e.g.\ Ghisellini et al.\ 1997; Raiteri et al.\ 2001; 
2003; Villata et al.\ 2002,  2009) determines the base level flux oscillations (Villata et al.\ 2004), while the medium-term 
can come from turbulence (Pollack et al.\ 2015).  Long term variations of BL Lacertae were found to be mildly
chromatic with slopes of $\sim$0.1 by Villata et al.\ (2004). We found a significant positive correlation between (V-R) against
R magnitude with a slope of 0.042, which can be explained by a Doppler factor variation on a spectrum slightly deviating from a 
power law shape and is convex for a bluer-when-brighter trend (Villata et al.\ 2004). Also, for the LTV, spectrum shape
 changes over time are plausible; due to them, the chromatism can also vary and the resulting superposition of different slopes could yield an overall 
flattening of the spectrum (Bonning et al.\ 2012).    \\

\begin{figure}
\centering
\includegraphics[width=8cm , angle=0]{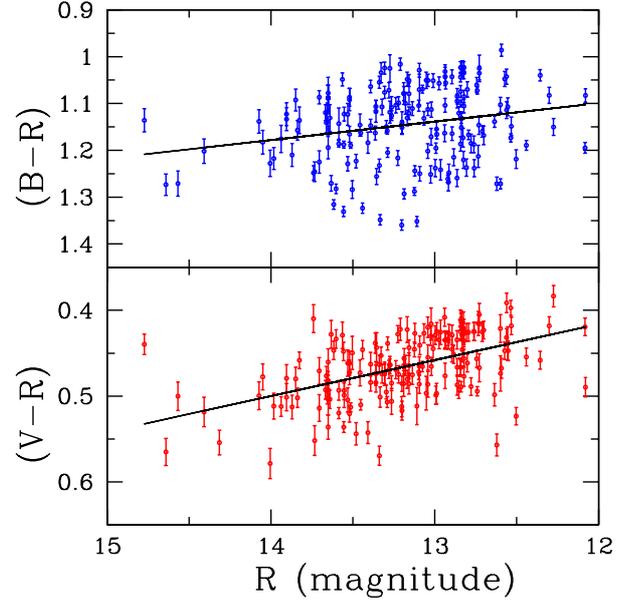}
\caption{Upper and lower panels respectively show the (B-R) and (V-R) behaviour with respect to R band magnitude.}
 \end{figure}

\subsection{Radio Data}
The radio flux density (in Jy) light curves on 36.8, 22.2, 14.5, 8 and 4.8 GHz are presented in the bottom panel of Figure 1.
Data at 230 and 86 GHz are taken from Raiteri et al.\ (2013) and are shown in the middle panel.
The most frequently sampled data were at 36.8 GHz; they show moderate activity until JD 2455750. After that one strong radio outburst
occured at around JD 2455890 and a more moderate one is seen at JD 2455960. It can be seen from the figure  that the
 light curves at all the radio frequencies exhibit similar behaviours and appear to be well correlated with each other 
during the first flare;  this is the case at 36.8 and 14.5 GHz during the second,
briefer, flare as well. However, at the radio frequency 22.2 GHz, there was a gap in the observations 
around JD 2455960 that is not clearly seen in the light curve as presented in Fig.\ 1, so that flare was
not detected. 

\begin{table*}
\centering
\caption{Variability analysis parameters at radio wavelengths.}
\begin{tabular}{llcccclcl} \hline
Freq. &      $<$S$>$    &  $m_0$  &m&   $Y$  &$\chi^2_r$&  $N $  &$\chi^2_{r~99.9\%}$& Telescope(s)  \\       
GHz(cm)  &      (Jy)       &(\%) & (\%)& (\%) &          &      &               &            \\\hline
4.8(6.25)  &4.30   &1.00 &1.63 & 5.73  & 6.362  &48   & 1.76          &UMRAO  \\
8.0(3.75)  &5.34   &1.00 &1.69 & 5.88 & 3.423   &64   & 1.64          &UMRAO  \\
14.5(2.07) &5.23   &1.00 &1.91 & 6.48 & 5.209   &124   & 1.44          &UMRAO       \\
22.2(1.35) &5.82   &0.45 &3.26 & 9.88 & 3.522    &107   & 1.48          &CrAO   \\
36.8(0.81) &5.56   &0.55 &3.60 & 10.91 & 2.416    &185   & 1.35          &Finland, CrAO   \\\hline
\end{tabular}  \\
$m$=variability index = $\sigma_{S}$/$<$S$>$,$\sigma_{S}$ standard deviation,   \\
$m_0$ = variability index of the secondary calibrators, \\
$Y= 3\sqrt{m^2 - m_0^2}$ = bias corrected variability
 amplitude (see Fuhrmann et al.\ 2008), \\
$\chi^2_r$= reduced Chi-square, \\
$N$ = number of data points, \\
$\chi^2_{r~99.9\%}$ = reduced Chi-square corresponding to a 
significance level of 99.9$\%$. \\
\end{table*}

 The variability parameters are characterized as in Fuhrmann et al.\ (2008).  
The mean flux $<$S$>$, variability index $m$, variability index of the calibrators $m_{0}$, noise 
bias corrected variability amplitude $Y$ and a reduced $\chi^2$ for a fit (where the BL Lac is considered to be 
variable if the $\chi^2$  gives a probability of $\leq$0.001 for the assumption of a constant flux) are presented in 
Table 1. \\
 
We find that BL Lacertae varies at all cm-band radio frequencies, on timescales of days with a variability index $m$ of about 1.6-3.6 per cent.
There is a systematic increase in the calibration-bias corrected variability index $Y$, from 5.73 at 4.8 GHz to
10.91 at 36.8 GHz.  This type of variability amplitude 
increase with frequency was also observed in earlier studies (e.g.\ Aller et al.\ 1985; Raiteri et al.\ 2003; Fuhrmann et al.\ 2008).
However, the opposite trend, i.e., variability amplitude decreasing with increase in frequency has also been observed 
(Beckert et al.\ 2002; Gupta et al.\ 2012).
These can be interpreted in terms of a variable source size arising from Interstellar
 Scintillation (ISS).  The type of change of variation strength with observed frequency thus can be used to find the  relative  dominance of ISS over 
source intrinsic variability (e.g.\ Beckert et al.\ 2002; Krichbaum et al.\ 2002; 
Gupta et al.\ 2012). The observed frequency dependence of the variability amplitude and the correlated variability 
across all bands in our observations argue in favour of a source instrinsic origin of these observed variations. 
These are usually explained in the blazars by synchrotron cooling and adiabatic expansion of a flaring component 
or a shock (e.g.\ Marscher \& Gear 1985).  \\

In order to consider the behaviour of the radio spectra, we first calculate the 36.8--14.5 GHz spectral index $\alpha$,
defined as $S_{\nu} \propto \nu^{\alpha}$,  for different brightnesses. 
 The behaviour of the spectral index is plotted with respect to the flux density at 14.5 GHz
and is shown in  Figure 4. There is a significant negative correlation with $r=-0.378$ and $p=0.0062$.
Hence, we see that the spectrum is becoming harder when the flux increases and we conclude that the strongly 
variable component dominates the radio spectrum then.\\

\begin{figure}
\includegraphics[width=6cm,angle=0]{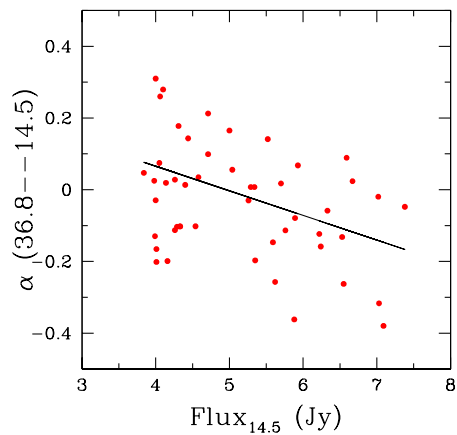}
\caption{Radio spectral index between 36.8 and 14.5 GHz against flux density at 14.5 GHz. }
\end{figure}

\begin{figure}
\includegraphics[width=6cm,angle=0]{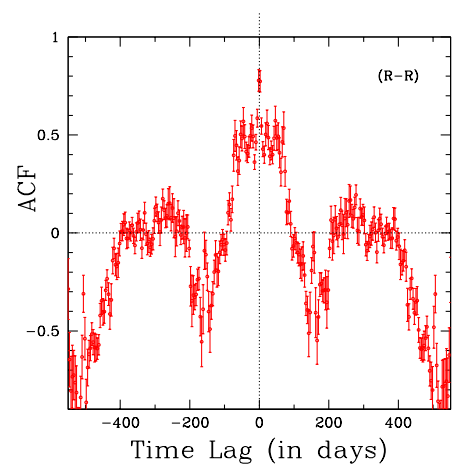}
\caption{ACF (R-R) of the optical data.}
\end{figure}

\begin{figure}
\includegraphics[width=6cm,angle=0]{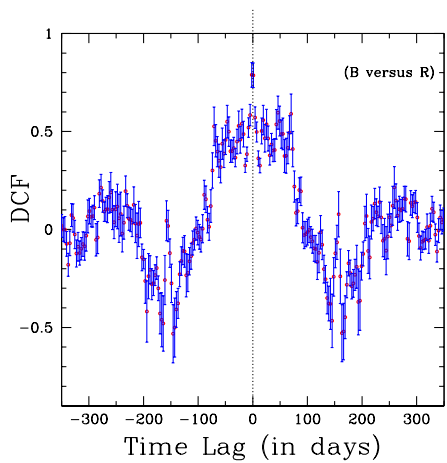}
\caption{DCF (B  vs.\ R) of the optical data.}
\end{figure}

\begin{figure}
\includegraphics[width=6cm,angle=0]{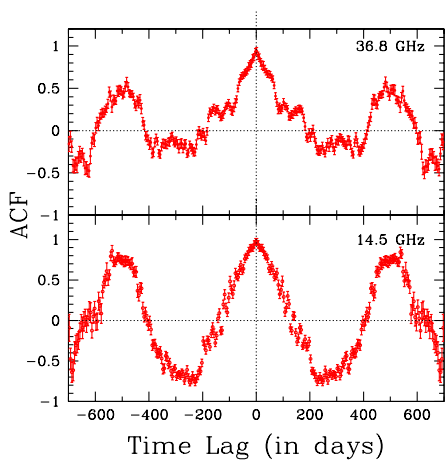}
\caption{ACFs of the radio data at 36.8 GHz (upper panel) and 14.5  GHz (lower panel).}
\end{figure}

\begin{figure}
\includegraphics[width=9cm,angle=0]{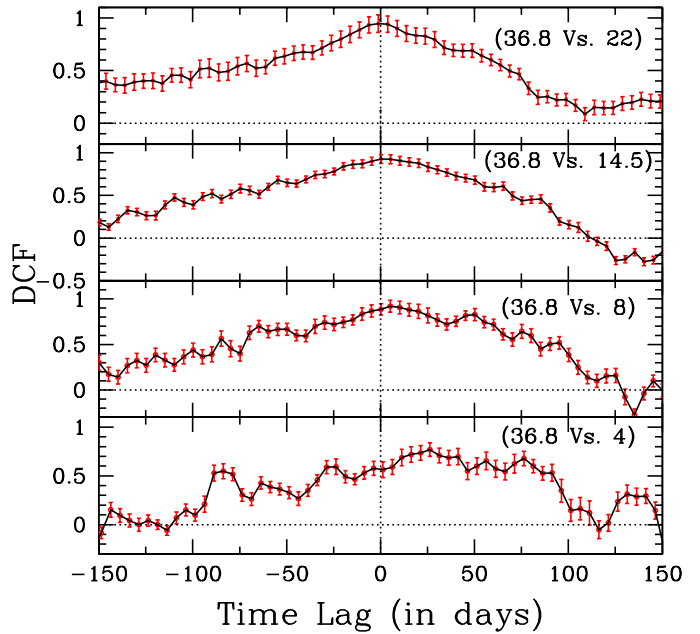}
\caption{DCFs of the radio data. Frequencies compared are written in each panel.}
\end{figure}

\subsection{Auto and Cross-correlations}
We computed Discrete Correlation Functions (DCF) following Edelson \& Krolik (1988) to search for 
possible variability time-scales and the time lags between multifrequency light curves, respectively. 
The first step is to calculate the unbinned correlation (UDCF) using the given time series by:
\begin{equation}
UDCF_{ij} = {\frac{(a_{i} - \bar{a})(b_{j} - \bar{b})}{\sqrt{\sigma_a^2 \sigma_b^2}}}.
\end{equation}
Here, $a_{i}$ and $b_{j}$ are the individual points in two time series $a$ and $b$, respectively, $\bar{a}$
and $\bar{b}$ are respectively the means of the time series, and $\sigma_a^2$ and $\sigma_b^2$ are their variances.
The correlation function is binned after calculation of the UDCF. 
The DCF can be calculated by averaging the
UDCF values for each time delay
\( \Delta t_{ij} = (t_{bj}-t_{ai}) \) lying in the range  \( \tau - \frac{\Delta\tau}{2} \leq t_{ij} 
\leq \tau+ \frac{\Delta\tau}{2} \) via

\begin{equation}
DCF(\tau) = {\frac{1}{n}} \sum ~UDCF_{ij}(\tau) .
\end{equation}
The  DCF analysis is used for finding the correlation and possible lags between the optical and radio frequencies.
 When the same data train is used, i.e., $a$=$b$, it is called the Auto Correlation Function (ACF) and has always has a  peak at zero lag, 
indicating that there is no time lag between the two, but any other strong peaks give indications of variability timescales.
The normalization of the DCF depends on the selection of the means $\bar{a}, \bar{b}$ and variances $\sigma_a^2$, 
$\sigma_b^2$, of each of the time series $a_{i}$ and $b_{j}$, respectively (White \& Peterson 1994). Hence, the most
accurate determination of the peak of the DCF depends on using the best estimates of the means and the variances of both of the 
series in Eqn.\ (1), which are generally taken to be the values determined from the entire time series. However, as these light curves 
are non-stationary statistical processes,  their means and variances change with time.
Hence, following White \& Peterson (1994), we computed the mean and variance for both the series  $a_{i}$ and $b_{j}$, using
only the points that fall within a given time-lag bin to contribute to the calculation of the DCFs at each $\tau$.

\subsubsection{Optical Correlations}
In order to find the variability timescales in the optical bands, we performed auto-correlations in various optical bands.
Figure 5 shows the ACF of the R band. The ACF of R band shows a peak at zero lag and then it decorrelates 
very fast with another peak at around 300 days. Since this peak is of very low significance, we do not claim 
 any variability timescale in the R band. Similar behaviours are shown in the ACFs of 
the light curves in the B, V and I bands hence we do not display them. We conclude that we did not find any characteristic variability 
time-scales in these optical measurements of BL Lacertae. \\

In order to examine any  time-delays between different optical bands, we performed a DCF between the B and R bands
which is shown in Figure 6. In the basic shock-in-jet model, time delays are expected between higher and lower frequencies
 as the high energy electrons emit synchrotron radiation first  and then cool, emitting at lower frequencies. 
The DCF between the light curves shows a peak of around 0.8, indicating close
correlation between the optical bands. In order to search for time delays, we fitted the peaks of the DCFs with
gaussian functions, but the time delay was consistent with zero lag. Similar behaviour was found with V versus R and I versus R
DCFs. Hence, we conclude that we did not find 
any optical inter-band time-lags on long-term trends.  This could be due to the small differences in frequency between the various optical bands.

\subsubsection{Radio Correlations}
Similarly, we performed auto-correlations for various radio frequencies to search for any characteristic timescales.
In Figure 7, we show the ACF of the 36.8 (upper panel) and 14.5 GHz (lower panel) frequency light curves in order to search for 
variability time-scales in radio bands. The ACFs of 36.8 and 14.5 GHz both show a strong peak (of $>$0.5) at $\sim$500 days time lag
but this timescales is just half of the duration of the overall coverage of the radio data, hence we do not regard this time-scale
to be significant. \\

It is clear from Figure 1 that all the radio frequencies exhibit  similar behaviours and that the flux variations are decreasing when
going from the higher to lower frequencies. In order to confirm this, 
we performed DCFs between the radio frequencies (36.8 vs. 22.2 GHz, 36.8 vs. 14.5 GHz, 36.8 vs. 8.0 GHz, and 36.8 vs. 4.8 GHz, shown
in top to bottom panel in Figure 8, respectively) and found that they are well correlated with each other in the sense that 
higher frequencies lead the lower frequencies. The DCF between 36.8 and 22.2 GHz frequencies (shown in the upper panel of 
Figure 8)  is very broad, but does not appear to be centered at zero lag;  we fit it with a gaussian function of the form:

\begin{equation}
DCF(\tau)=a \times {\rm exp}[\frac{-(\tau - m)^{2}}{2 \sigma^{2}}] .
\end{equation}

 Here, $a$ is the peak value of the DCF; $m$ is the time lag at which DCF peaks and $\sigma$ is the width of the Gaussian function.
The error in the location of the peak of the DCF is clearly much smaller than the width of the Gaussian function
 (e.g.\ Peterson et al.\ 1988). We found that the variations at higher frequencies are very well correlated 
with a lag of $2.11\pm1.19$ days, with 36.8 leading the 22.2 GHz. By visual inspection, it is clear that lower frequencies 
are lagging the higher frequencies in other panels too.  The gaussian function fits to the  DCFs give  time lags 
of $\sim 2.90 \pm 0.66$ days between  36.8 GHz and  14.5 GHz (second panel), $\sim 8.86 \pm 2.71$ days between 36.8 and 
8 GHz (third panel) and $\sim 25.76 \pm 7.25$ days between 36.8 and 4.8 GHz
(bottom panel), respectively. Hence, the flux variations at the higher frequencies are leading with 
respect to the lower frequencies, as is very common in blazars (e.g. Marscher \& Gear 1985; Aller et al.\ 1999b;
 Raiteri et al.\ 2001; Raiteri et al.\ 2003; Gupta et al.\ 2012). \\

Flux variations where shorter wavelengths lead the longer ones are commonly interpreted as a source intrinsic opacity effect
due to synchrotron self-absorption (e.g., Kudryavtseva et al.\ 2011) which is further related to the source activity and the 
shape of the radio spectrum. Similar evidence was found by Gupta et al.\ (2012),
where they detected a significant time lag between 2.8 cm and the longer wavelengths, again with shorter wavelengths leading the longer
ones. \\

\subsubsection{Optical versus Radio Correlations}

The optical and radio light curves of our observations are presented in Figure 1. BL Lac exhibited multiple flares at optical frequencies
in all  bands. We have included the published WEBT data of optical R band of Raiteri et al.\ (2013) to make the data sampling even better. To 
investigate the possible correlations between optical and radio frequencies, we performed DCF analysis for the almost three-year
long simultaneous data trains. There are two prominent flares in radio frequencies, one at JD 2455890 and the other at JD
2455960 and we first performed cross-correlations of the optical R band with respect to the radio frequency at 36.8 GHz; 
 which is the best sampled one, given that, as noted above, the second flare was missed at 22.2 GHz.  
  The light curves reveal that the behaviour of the radio frequencies are quite different from the optical ones and 
there is no immediate correspondence between the flares in optical bands and the radio ones.\\

 The DCF between R and 36.8 GHz band for the whole observing season
 is shown in Figure 9 and produces a peak at around 250.28$\pm$10.21 days. We fit the 
peak with gaussian functions of the form described above and found significant correlations between them
where positive lags indicate that the R band is leading the radio frequency.  
The peak in the DCF presumably arise predominantly from connecting the optical outburst at around JD 2455700 with the  
radio outburst around JD 2455960. In the DCF, there is no significant
correlation between R band and radio flux at zero time lag which rules out the possibility of simultaneous optical and radio flares.
We also performed cross-correlations between optical R band and the lower radio frequencies in order to investigate the delays between them;
we fit their DCFs with  Gaussians and found peaks between 235 and 260 days.\\

The lags of around 235--260 days of the cm-frequencies with respect to the optical 
ones are  in reasonable accordance with the results of Villata et al.\ (2009) for BL Lacertae during the period 1994--2003 
where they found fair optical--radio correlations with 
different time delays of optical with respect to the radio events varying from about 100 days to 300 days.
The variable radio delay could be explained by the jet changing its orientation with 
respect to the line of sight, in that relativistic effects
produce a variation in the observed time scales;  if the jet region emitting the optical-to-radio radiation makes a smaller viewing
angle, it yields a shorter delay of the radio events with respect to the optical ones (e.g.\ Villata et al.\ 2009).  \\

\begin{figure}
\includegraphics[width=8cm,angle=0]{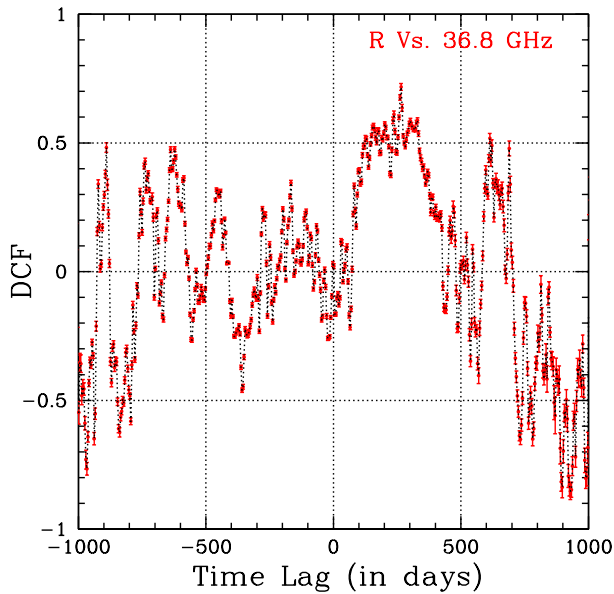}
\caption{DCF between optical R band versus radio frequency at 36.8 GHz.}
\end{figure}

\section{Discussion \& Conclusions}
We performed observations of the well known blazar
BL Lacertae in optical and radio bands during the period 2010--2013 and 
searched for possible correlations among them. Our optical observations were obtained from seven ground based telescopes
and taken on  almost 200 nights. The source was very active and showed repeated `mini-flares' of 0.2--0.5 magnitude in amplitude
 on short timescales. Colour analysis of the long term light curve reveals a significant correlation between the 
colour index and the source brightness, with a bluer-when-brighter trend. But the slope of this trend is only 0.042, which is mildly chromatic 
according to Villata et al.\ (2004).  Hence it can be explained by the superposition of different slopes during the flaring states,
leading to an overall flattening of the spectrum.\\

Radio light curves at different frequencies are well correlated, with the flux amplitude becoming smaller with a decrease in radio frequency.
The radio spectral index decreases with source brightness, which is very commonly  observed in blazars.
We cross correlated the different radio frequencies and found that flux variations at lower frequencies lag those at higher frequency
with time delays ranging from days to weeks. This can be interpreted as the higher frequencies being emitted from the inner and denser
parts of the jets, with the lower frequencies being emitted from comparatively more distant and less dense parts. Alternatively, Villata \& Raiteri (1999)
explained this type of behaviour by a rotating helical jet model which causes time lags since the different frequency emitting 
portions of the jet acquire the same viewing angle at different times. \\

Many authors have studied the correlations between optical and radio bands in the literature but the results of different sources
do not show a common behaviour. On IDV timescales, radio and optical bands have been found to be correlated  for the blazar 
S5 0716+714, which strongly favours an intrinsic origin for the former instead of the ISS effect (e.g., Quirrenbach et al.\ 
1991; Wagner et al.\ 1996; Fuhrmann et al.\ 2008 and references therein).
However, Gupta et al.\ (2012) did not find any significant correlation between these bands for the same source, and attributed the
the lack of correlation at that time to different physical causes: ISS at long radio wavelengths, but intrinsic origin in the optical. \\

 On longer timescales, Villata et al.\ (2009) found significant delays among variations in optical and radio bands by about 100 days 
in the WEBT campaign of BL Lacertae and  explained them in terms of an emitting plasma flowing along a rotating helical path in a curved jet.
On other occasions, very weak correlations are found for BL Lac between optical and radio bands (i.e., B{\"o}ttcher et al.\ 2003; 
Raiteri et al.\ 2003; Guo et al.\ 2015, and references therein).  \\

In searching for the possible correlations and time lags in the optical and radio band for the observing seasons 2010--2013
where BL Lacertae has shown outbursts in both the bands,
 we included the published data of the optical R band from Raiteri et al.\ (2013) to get better data sampling. 
 Major optical outbursts, such as the strong flare at JD 2455700, do not produce simultaneous radio flux enhancements 
but strong radio flares at JD 2455890 and 2455970 appear to be associated with the  optical flares with time delays. 
In order to investigate the correlations, we performed DCFs between the R band and various radio frequencies 
and found significant correlations between them with optical leading the radio frequencies. The typical time lag  
of $\sim 250$ days  between the optical and the various radio frequencies
 appears to be due to the connection of the strong optical outburst at JD 2455700 with the strong radio outburst 
at JD 2455960.  Raiteri et al.\ (2013) performed
the correlation between their R band and at mm-flux densities (at 230 and 86 GHz) and found strong correlation at a lag of 120--150
days. Hence, our results are consistent with those of Raiteri et al.\ (2013) since higher radio frequencies usually show
lesser delays with respect to the optical emission. The delayed cm radio lags with respect 
to the R band are in accordance with the inhomogeneous jet model, where synchrotron photons at lower frequencies are emitted from 
progressively more external regions as compared to the higher frequency photons.

In previous studies of BL Lacerate, it has been argued that the long term variability arises due to the intrinsic as well as geometrical
effects and the relativistic effects produce a variation in the observed timescales due to change of the jet orientation with respect
to our line of sight (Villata et al.\ 2009; Raiteri et al.\ 2013, and references therein). During this period of 2010--2013, 
the behaviour of BL Lacerate can be explained by taking into account these scenarios where the higher radio frequencies lag behind
the optical by about 120--150 days  and then the mechanism producing the flux enhancements propagates downstream
 where it produces the lower frequencies which lag the optical  $\sim$250 days. Much longer and better sampled light curves are required to
better understand the correlation and time lags between these bands.\\

\begin{acknowledgements}
 We thank the referee for useful and constructive comments.  We thank Dr.\ C.~M.\ Raiteri for providing us with
the WEBT R-band data. This work is partly based on data taken and assembled by the WEBT collaboration and stored in the WEBT 
archive at the Osservatorio Astrofisico di Torino - INAF (http://www.oato.inaf.it/blazars/webt/). 
This research was partially supported by Scientific Research Fund of the Bulgarian Ministry of Education and Sciences under
grant DO 02-137 (BIn 13/09). The Skinakas Observatory is a collaborative project of the University of Crete, the Foundation
for Research and Technology -- Hellas, and the Max-Planck-Institut f\"ur Extraterrestrische Physik.
H.G.\ is sponsored by the Chinese Academy of Sciences Visiting Fellowship for Researchers from Developing Countries
 (grant No.\ 2014FFJB0005), and supported by the NSFC Research Fund for International Young Scientists (grant No.\ 11450110398).
A.C.G.\ is partially supported by the Chinese Academy of Sciences Visiting Fellowship for Researchers
from Developing Countries (grant no. 2014FFJA0004). 
 M-F.G.\ acknowledges support from the National Science Foundation of China (grant 11473054) and the
Science and Technology Commission of Shanghai Municipality (14ZR1447100). 
The Abastumani team acknowledges financial support of the project FR/639/6-320/12 
by the Shota Rustaveli National Science Foundation under contract 31/76.
The work at UMRAO was supported in part by a series of grants from the NSF and from the NASA 
Fermi Guest Investigator Program.
The work is partially supported by India-Ukraine inter-governmental project
‘Multiwavelength Observations of Blazars’, no. INT/UKR/2012/P-02.
The Mets\"ahovi team acknowledges the support from the Academy of Finland to our observing projects
 (numbers 212656, 210338, 121148, and others). This paper is partly based on observations carried out at 
the IRAM 30-m Telescope, which is supported by INSU/CNRS (France), MPG (Germany) and IGN (Spain).
\end{acknowledgements}

\end{document}